\documentclass[12pt,a41]{article}

\usepackage{epsfig}

\begin{document}
\begin{center}
{\LARGE\bf Results on $\Lambda^0$ Production\\[2mm] at HERMES}

\vspace{.5cm}
{P. Chumney representing the HERMES collaboration.}\\
{\it New Mexico State University, Las Cruces, NM 88003, USA}\\
\vspace*{.5cm}
\end{center}

\begin{abstract}
The production of $\Lambda^0$'s at the HERMES experiment
is presented.
Prospects for the future of $\Lambda$ measurements at HERMES
are discussed.
\end{abstract}

\section{Introduction}

The HERMES Experiment was designed primarily to study the spin 
structure of the nucleon in polarized deep inelastic electron/positron
nucleon scattering.
It uses the polarized positron beam at HERA and an internal gas target.
For more details, see the other HERMES contributions to these 
proceedings, Ref. \cite{Hasc97,Bron97,Geig97,Tall97}

The strength of HERMES lies in its ability to detect the scattered positron 
and 
the hadronic final state.
At HERMES many particles such 
as $\Lambda^0$'s can be reconstructed from the final state hadrons in numbers
that make detailed analysis of their properties 
possible.  Since the $\Lambda$ decay 
violates parity, the polarization can be
reconstructed from the angular distribution of the decay products.
At HERMES the target and beam polarization will affect the 
$\Lambda$ polarization, either by transfer of polarization from the
nucleon's quarks, or through the polarized virtual photon. 
HERMES can detect the current and target  
fragmentation region of $\Lambda$ production.  The current 
fragmentation will give information about the $u$ quark fragmentation and
the target fragmentation will provide insight into the polarization of
the strange sea.

\section{Lambda Observables}
\subsection{Current Fragmentation}

In the current fragmentation region ($x_f>0$, $x_f$ is the fraction
of available longitudinal momentum, defined to be along the beam line,
 transferred to the hadron 
in the $\gamma^*N$ center of mass frame) $\Lambda$'s acquire their
polarization during the polarized quark fragmentation.    
ALEPH \cite{Busk96} and OPAL\cite{Acke97} have recent significant negative 
results for longitudinal polarization from $Z^0$ 
decays in $e^+e^-$ collisions.  This is due to the fragmentation of 
$s$ quarks which are highly polarized \cite{Busk96}.  For deep inelastic
scattering, the $u$ is the dominant quark \cite{Jaff96} and a substantial
longitudinal $\Lambda$ polarization would be a signature for spin transfer
of the virtual photon to the $u$ quark and correspondingly of the 
$u$-quark polarization in the $\Lambda$.

For a positron or electron beam of polarization $P_B$
on an unpolarized target the resulting $\Lambda$
polarization is predicted to be (for a small positron 
scattering angle)\cite{Jaff96}:
\[
P_\Lambda^L = P_B D_y \tau_B=P_B \frac{1-(1-y)^2}{1+(1-y)^2} \tau_B=
P_B \frac{1-(1-y)^2}{1+(1-y)^2}\frac{\sum_qe^2_qq_N(x)
\Delta D^\Lambda_q(z)}{\sum_qe^2_q q_N(x) D^\Lambda_q(z)}
\]
where $D_y$ is the depolarization factor, 
$D_y\tau_B$ is the polarization transferred from the beam positrons to
the $\Lambda$, $e_q$ is the charge for quark $q$,
$q_N(x)$ is the spin average
quark distribution for the nucleon, 
$\Delta D^\Lambda_q(z)$ is the spin dependent 
fragmentation function for a quark fragmenting into a $\Lambda$, 
and $D^\Lambda_q(z)$ are the corresponding spin-independent fragmentation 
functions.  Information about the u-quark fragmentation function ratio
$\Delta D^\Lambda_u(z)/D^\Lambda_u(z)$  in the current fragmentation 
region can be obtained.  Estimates of precision for an experiment like HERMES 
based on the $u$ quark dominance have been calculated elsewhere \cite{Jaff96}. 

\subsection{Target Fragmentation}
In the target fragmentation region ($x_f<0$) the struck quark has non-zero
net longitudinal polarization, and Ellis' model \cite{Elli96}
suggests that the remnant $s$ quark will have a {\it negatively-correlated}
polarization which is transferred to the $\Lambda$. 
The polarization in the target fragmentation 
region is predicted to be \cite{Elli96}:
\[
P_\Lambda^L=\frac{\sum_qe^2_q[P_BD_yq(x)+P_T\Delta q(x)]c_{sq}}
{\sum_qe^2_q[q(x)+P_B P_TD_y\Delta q(x)]}
\]
$\Delta q(x)$ is the helicity difference quark distribution, $D_y$ is 
the same depolarization factor as for current fragmentation, 
$c_{sq}$ is the spin-correlation coefficient:
\[
c_{sq}=\frac{P_{s+q} - P_{s-q}}{P_{s+q} + P_{s-q}}
\]
and $P_{s+q}$($P_{s-q}$) represents the probability that the spin of the 
remnant $s$ quark is parallel(antiparallel) to the spin of the struck quark.
Measurement of $P_\Lambda$ in the target fragmentation region
will give insight into the polarization of
the strange sea quarks.  

\subsection{Measurement}
Since the $\Lambda\rightarrow p\pi$ decay
violates parity, its polarization can be determined by the angular
distribution of its decay products.
To measure longitudinal $\Lambda$ polarization, the cosine of the 
angle between the direction of the $\Lambda$ momentum in
the lab and the proton ($\cos\theta_0$) in the $\Lambda$ rest frame is
reconstructed and the mean value calculated ($\langle\cos\theta_0\rangle$).  
However, the acceptance of the HERMES
spectrometer is asymmetric in $\cos\theta_0$.  This is because the pion's 
momentum is much softer than that of the proton, and the pion is more likely
to miss all or part of the HERMES acceptance.
It is necessary, therefore, to reverse the beam polarization and compare  
$\cos\theta_0$ distributions for both helicities to obtain the polarization.

\section{Data}

HERMES had two periods of dedicated unpolarized target 
running in 1996 and 1997 between which the beam helicity was
reversed. 
An unpolarized target allows for a higher density  
(${Luminosity}\sim10^{32}$~cm$^{-2}$~s$^-{1}$) than the polarized target,
so that a large number of deep inelastic events can be collected over
a short period of time.
In addition, the $\Lambda$
analysis becomes simplified since it does not have to
take into account the polarization transfer from the target nucleons.
A short summary of the events collected for the two periods of running is
shown in Table~\ref{tab:unpol}.

\begin{table}[h]
  \begin{center}
  \leavevmode
  \begin{tabular}{lll}
  Year:&1996&1997\\
  \hline
  Target Types:&H,D,$^3$He&H,D,N\\
  Number of DIS:&5.4$\times10^6$&4.5$\times10^6$\\
  Number of Lambdas:&3960&4711\\
  With DIS Cuts: &2695&2609\\
  \end{tabular}\\[.5cm]   
  \caption{Summary of HERMES statistics collected during 
unpolarized target running.}
  \label{tab:unpol}
  \end{center}
\end{table}

\begin{figure}[h]
  \begin{center}
    \leavevmode
 \epsfig{file=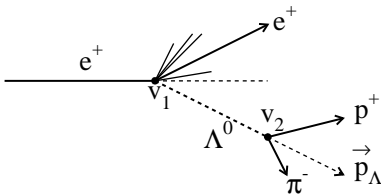, height=2.75cm, width=5cm}\\[.5cm]   
    \caption{A simplified diagram of a $\Lambda$ event, showing how 
the different vertices are defined.}
    \label{fig:lamevt}
  \end{center}
\end{figure}

In each candidate $\Lambda$ event:
\[
e\:+\:N\:\rightarrow\:e^\prime\:+\Lambda(\bar{\Lambda})\:+\:X
\]
HERMES identifies a positron and two hadrons from the $\Lambda$ decay. 
Momenta $\vec{p}_{e^\prime},\:\vec{p}_p,\:\vec{p}_{\pi}$ are measured, 
vertices $v_1$, $v_2$, and invariant mass 
$M_{p\pi}$ are reconstructed.  How the vertices
 are defined is shown in
Figure~\ref{fig:lamevt}.   After reconstruction
of the event, several basic requirements must be met for the event
to be considered a $\Lambda$ event.
There must be at least three tracks in the detector, one track a positron, 
and a decay vertex with two oppositely charged hadron tracks. In addition, 
deep inelastic scattering (DIS) cuts are made on the data:  
negative four-momentum transfer of the lepton $Q^2\:>\:1.0\:{\rm GeV}^2$, 
invariant mass of the hadronic system $W^2\:>\:4.0\:{\rm GeV}^2$, 
fractional energy transfer of the lepton $y\:<\:0.85$, and the 
momentum fraction of the struck quark $0.02\:<\:x_{Bj}\:<\:0.8$ \cite{Belo97}.

\begin{figure}[h]
\begin{center}
\epsfig{figure=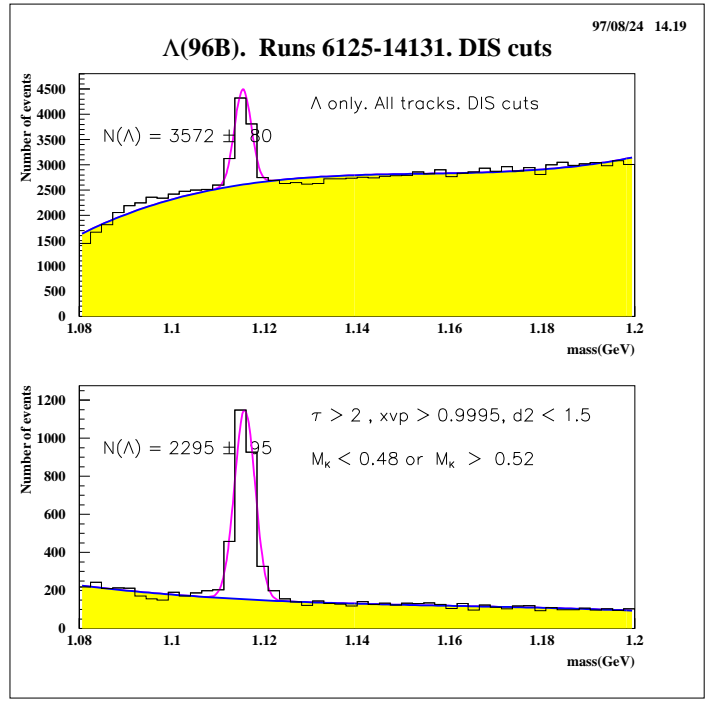, width=9.25cm}
\caption{1996 $\Lambda$ invariant mass distributions for two different sets of cuts. The upper portion is with DIS cuts only, and the lower plot has the
additional cuts imposed.   Both plots are for the entire 1996 running.  
The reduction of the background is significant.
(HERMES detects about 1/5 as many $\overline{\Lambda}^0$'s.)}
\label{fig:dis96}
\end{center}
\end{figure}

After all these requirements, however, there still remains a substantial
background.  This can be significantly reduced by requiring a set of
tighter cuts which take into account the properties of
$\Lambda$ decays. These include: the direction of $\Lambda$ momentum 
($\vec{p}_\Lambda$) 
should be close to direction ($\vec{v}$) defined by primary $e^+$ ($v_1$) and 
secondary $p\pi$ ($v_2$) vertices:
\[
x_{vp} = \frac{(\vec{v},\vec{p}_\Lambda)}{|\vec{v}|\cdot|\vec{p}_\Lambda|}>0.9995,
\]
the separation of tracks $d_2$ at the secondary $\Lambda$ vertex $v_2$
 should be small:
\[
d_2 < 1.5 {\rm cm},
\]
the calculated decay length of the $\Lambda$ should be:
\[
c\tau > 2\:{\rm cm},
\]
and a requirement that the invariant mass of the decay vertex on
the assumption of a $\pi^+\pi^-$ instead of $p^+\pi^-$ does
not give a $K_s^0$ mass:
\[
M_{\pi\pi} < 0.48\:{\rm or}\:M_{\pi\pi}>0.52.  
\]
The effects of the basic cuts and the stricter cuts is shown for 1996 data
 in Figure~\ref{fig:dis96} \cite{Belo97}.  The reduction of the background
can clearly be seen. 

\subsection{Event Types}
\begin{figure}[h]
  \begin{center}
    \leavevmode
    \epsfig{figure=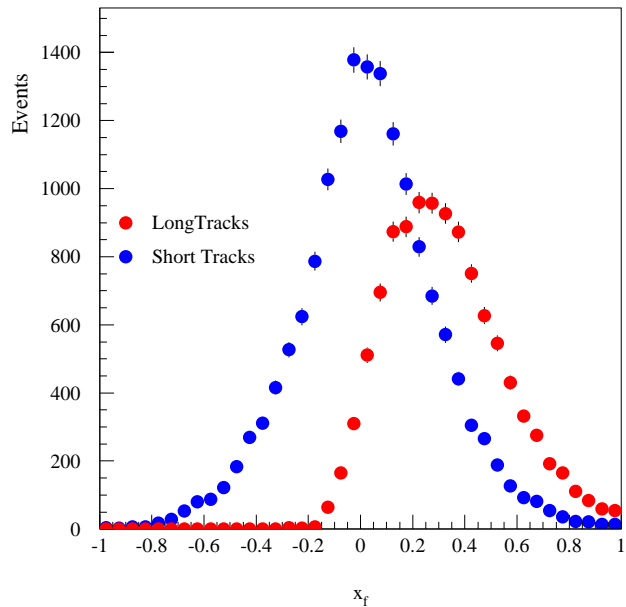,height=9.5cm,angle=-90.}
    \caption{The different $x_f$ distribution for short and long track events.
The short track events are almost symmetric about $x_f=0$ and are well
within the target fragmentation region.  The long tracks are mostly at 
$x_f>0$ and in the current fragmentation region.}
    \label{fig:xfsl}
  \end{center}
\end{figure}

Figure~\ref{fig:xfsl} shows two types
of HERMES events. HERMES can reconstruct tracks where
the particle may leave the detector acceptance before it reaches the
calorimeter, these are called short tracks.  However the momentum
reconstruction is not as good because the there is not a track in 
the rear tracking chambers of the detector.  The track's bend in the magnet
only goes through the magnet proportional chambers and leaves
before it is tracked in the rear tracking chambers.
In both kinds of events 
the positron always has a full track through the detector.
A long track event is when the $\pi^-$ and $p$ have full tracks in HERMES.
A short track event is when either the $\pi^-$, $p$, or both particles have 
tracks only recorded in the front region of HERMES.

\begin{figure}[h]
  \begin{center}
    \leavevmode
\epsfig{figure=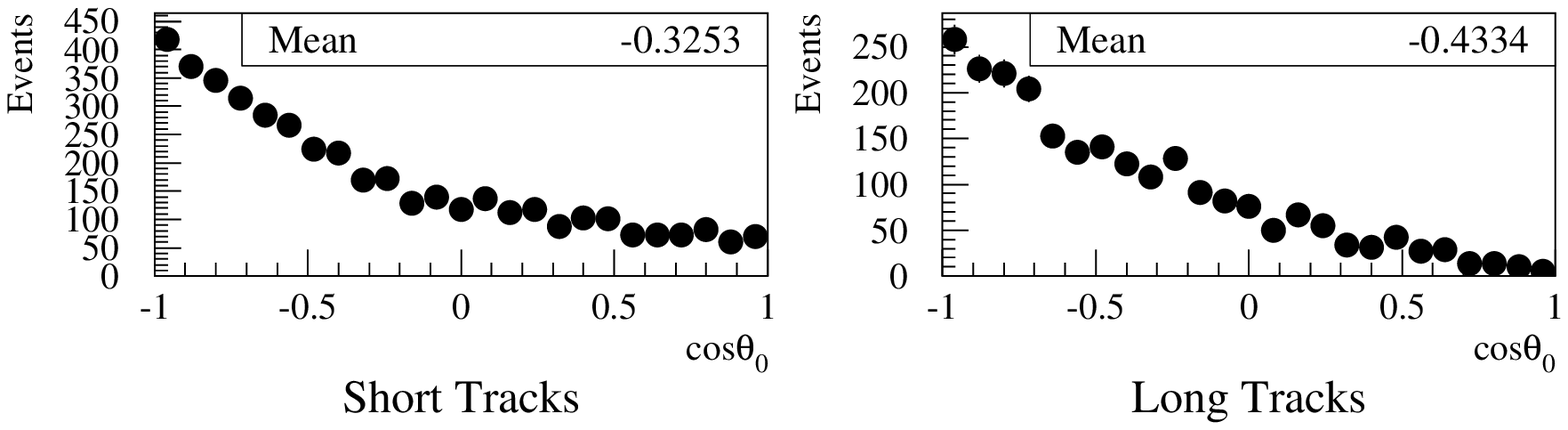,height=5.0cm}
    \caption{Monte Carlo $\cos\theta_0$ acceptance for short and long track events.  The asymmetry from the HERMES acceptance reduces by 25\% for the short track events in this group of MC events.}
    \label{fig:cossl}
  \end{center}
\end{figure}

The short track events allow access to the target
fragmentation region, shown for data in Figure~\ref{fig:xfsl}.  In addition,
the asymmetric $\cos\theta_0$ distribution, caused by the
experimental acceptance, is reduced 
as can be seen in Figure~\ref{fig:cossl} for Monte Carlo unpolarized 
$\Lambda$'s.

\section{Monte Carlo}

An integral part of the study of $\Lambda$ physics at HERMES has been
understanding the parent particle distributions 
and the experimental acceptance with the help
of Monte Carlo simulation.  For example, the 
$\Sigma^0$'s can depolarize the resulting $\Lambda$ about 
30\% of the time \cite{Jaff96,Burk93}.
Monte Carlo results, summarized in
Table~\ref{tab:MCpar}, show that only 16\% are expected
from this channel, resulting in a 5\% correction of the
measured total $\Lambda$ polarization.  
\begin{table}[h]
\begin{center}
\begin{tabular}{|l|r|r|}
\hline
Parent & Decay Mode & Percentage \\
\hline
$\Sigma (1385)$ & $\Lambda\pi$ & 30\% \\
$\Sigma^0$ & $\Lambda\gamma$ & 16\% \\
$\Xi^0,\Xi^- $ & $\Lambda\pi^0,\Lambda\pi^-$ & 5\% \\
\hline
\end{tabular}
\caption{Monte Carlo parent particle fractions. The remaining 49\% are
produced by string fragmentation.}
\label{tab:MCpar}
\end{center}
\end{table}

\begin{figure}[h]
\begin{center}
\leavevmode
\epsfig{figure=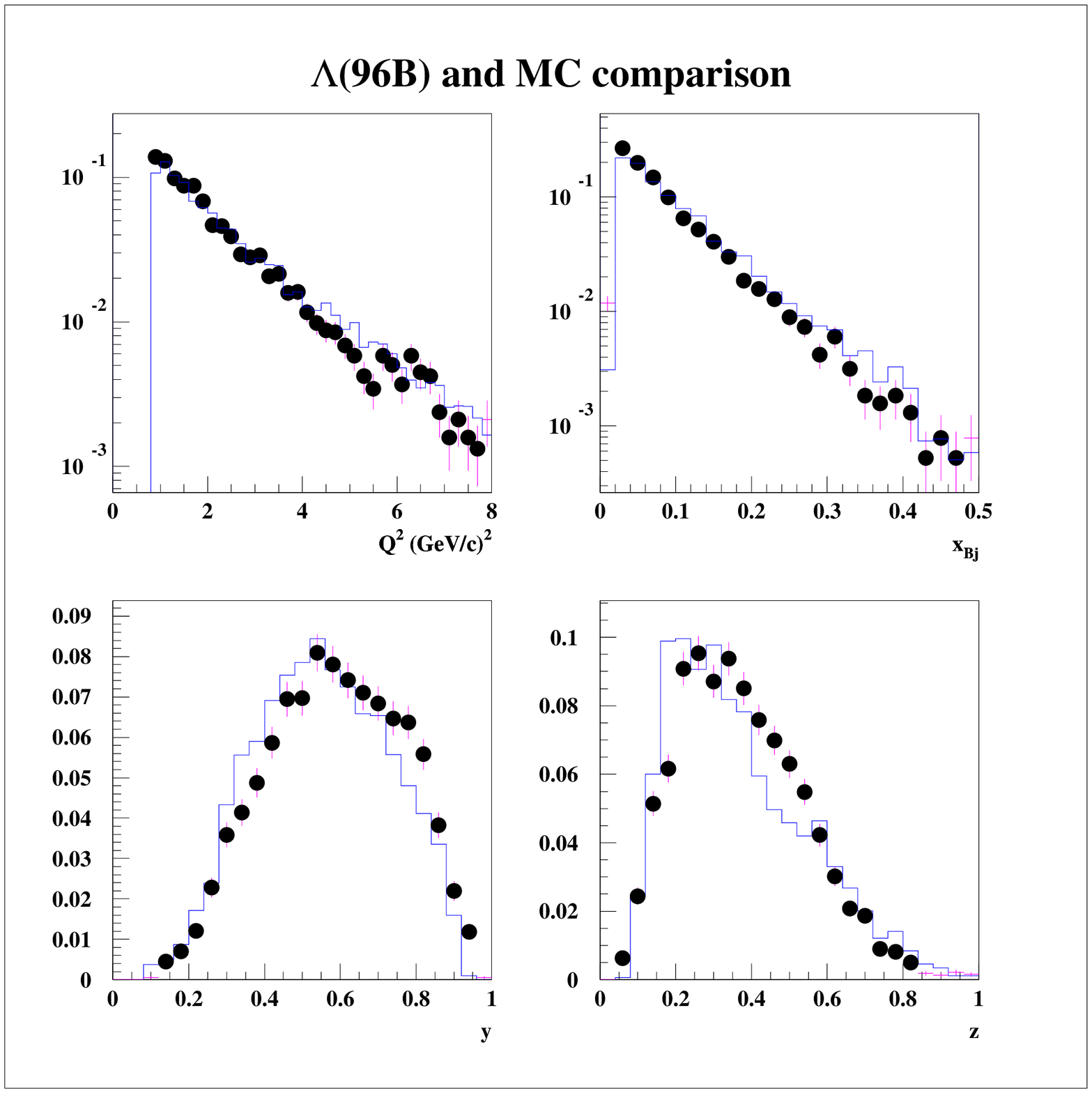,width=9.25cm}      
\caption{Comparison of kinematic distributions for data (black circles) and MC (lines) normalized to the total number of events in both data sets. There is 
reasonable agreement between data and MC.}
\label{fig:MCKine}
\end{center}
\end{figure}

In addition, kinematic distributions must be compared to make sure the
$\Lambda$ Monte Carlo reproduces the data.  In Figure~\ref{fig:MCKine} the 
distributions,
after DIS cuts, are shown for $Q^2$, $x_{Bj}$, $y$, and the Lorentz
scaling variable $z=\frac{E_\Lambda}{\nu}$ 
which relates $E_\Lambda$ to the
energy of the virtual photon.  
The Monte Carlo and data show reasonable agreement
with some subtle differences.  

\section{Future Plans}
\subsection{Lambda Wheels}
Currently there is a
proposal to put two round silicon detectors, called Lambda Wheels (LW) 
downstream of the target before $1999$ running \cite{Amar97}. They will 
be mounted at about 50 cm from the center of the target. Each LW will be 
two layers with a total thickness of 300~$\mu$m silicon each. 
The inner radius will be 7~cm and the outer radius 17.5~cm, 
and an opening angle of 340~mrad.  This will 
increase the detection of mostly large-angle pions which leave the
HERMES acceptance.  Protons will still be detected in the main
HERMES spectrometer.
 
Simulation was done to test the effectiveness of adding the Lambda Wheels.
For this, 32584 unpolarized $\Lambda$ events were generated.
Improvements in the number of events detected, mean $x_f$,
 and $\langle\cos\theta_0\rangle$ are summarized 
in Table~\ref{tab:LWMC}.  
\begin{table}[h]
\begin{center}
\leavevmode
\begin{tabular}{lrrr}
Tracks: & Mean $x_f$ & $\langle\cos\theta_0\rangle$ 
& N$_{events}$ \\
\hline
Long & 0.1325 & -0.341 & 4100 \\
Short+Long & 0.0811 & -0.184& 8300 \\
LW+Short+Long & -0.0287 & 0.004 & 17100 \\
\end{tabular}
\caption{ $x_f$ and $\langle\cos\theta_0\rangle$ for the events based
on their track and detection characteristics.}
\label{tab:LWMC}
\end{center}
\end{table}
Overall there is a factor of 4 improvement in statistics, 
a larger fraction of events are in the target 
fragmentation region, and a very good reduction 
of the acceptance-caused asymmetry in the $\cos\theta_0$ 
angular distribution.

\subsection{Transversely Polarized Target}
It is also proposed to have a transversely polarized target.
The transverse $\Lambda$ polarization for a transversely polarized target 
and unpolarized electron beam
is predicted to be\cite{Jaff96}:
\[
P_\Lambda^T=\frac{2(1-y)}{1+(1-y)^2}
\frac{\sum_q\delta q_N(x,Q^2)\delta D_\Lambda(z,Q^2)}
{\sum_q q_N(x,Q^2) D_\Lambda(z,Q^2)}
\]
The $\delta q_N(x)$ are the transversity difference quark
distributions in the nucleon
and the $\delta D_\Lambda(z,Q^2)$ are the $\Lambda$ transverse 
fragmentation functions.
Using a transverse target will allow us to measure transverse 
$\Lambda$ polarization $P_\Lambda$.
In addition, it might be  possible to measure 
the transverse structure function 
$h_1(x,Q^2)$ via $\Lambda$ asymmetries\cite{Artr90,Coll94,Jaff96-2}.

\section{Conclusions}
HERMES will extract a preliminary longitudinal $P_\Lambda$ from 
1996 and 1997 data in the near future.  
The expected statistical accuracy for $P_\Lambda$
is about 0.013 for 
$\sim2600$ $\Lambda$'s in each helicity data set. 
Continuing in 1999, HERMES will increase 
the $\Lambda$ statistics and explore target fragmentation with 
the Lambda Wheels. The possibility of running with a  
transversely polarized target 
promises more exploration of interesting $\Lambda$ physics.

\vspace{1mm}
\noindent

\vspace{1mm}
\noindent
%\bibliography{pams}

\end{document}